\documentclass[5pt, 11pt]{elsarticle}
\usepackage{amssymb}
\usepackage{fullpage}
\journal{Nuclear Physics B Proceedings Supplement}

\begin{document}
\begin{frontmatter}
\title{PROSPECT - A precision oscillation and spectrum experiment}
\author{T.J. Langford \textit{for the PROSPECT Collaboration} \\ \textit{Wright Laboratory - Department of Physics - Yale University \\ New Haven CT 06511}}

\begin{abstract}
Segmented antineutrino detectors placed near a compact research reactor provide an excellent opportunity to probe short-baseline neutrino oscillations and precisely measure the reactor antineutrino spectrum.
Close proximity to a reactor combined with minimal overburden yield a high background environment that must be managed through shielding and detector technology.
PROSPECT is a new experimental effort to detect reactor antineutrinos from the High Flux Isotope Reactor (HFIR) at Oak Ridge National Laboratory, managed by UT Battelle for the U.S. Department of Energy.  
The detector will use novel lithium-loaded liquid scintillator capable of neutron/gamma pulse shape discrimination and neutron capture tagging.
These enhancements improve the ability to identify neutrino inverse-beta decays and reject background events in analysis.
Results from these efforts will be covered along with their implications for an oscillation search and a precision spectrum measurement.
\end{abstract}


\end{frontmatter}

\section{Introduction}\label{sec-intro}

The recent precision measurements of $\theta_{13}$ using identical detectors placed near and far from PWR-style power reactors have demonstrated the importance of reactor antineutrino experiments~\cite{An:2012eh,Abe:2012tg,reno:2012}.
Over the course of three years, these experiments have taken an unknown mixing angle and made it the most precisely measured value of all three. 
However, two interesting questions have arisen from these experiments. 
First is the survival of the reactor anomaly despite the impressive statistical and systematic uncertainties in these experiments. 
Second is the deviation from the predicted neutrino spectrum between 5 and 6~MeV that has now been observed by all three experiments. 
The appearance of the spectral deviation has called into question uncertainties in the predicted flux of antineutrinos from reactors and could be a key to understanding the reactor anomaly. 
Recent work has suggested that the spectral deviation could be explained by performing an \textit{ab-initio} calculation of the spectrum rather than relying on beta conversion measurements~\cite{dwyer2014spectral}.

A new experiment has been proposed that is able to directly address both of these questions.  
The PROSPECT experiment consists of a segmented liquid organic scintillator detector located at the High Flux Isotope Reactor at Oak Ridge National Laboratory, managed by UT Battelle for the U.S. Department of Energy~\cite{xoubi2005modeling, Ashenfelter2013}.
Split into two phases, the experiment will consist of a near detector (phase-1) and a larger far detector (phase-2).
The phase-1 detector consists of 140 identical segments that allow for precise distance determination and enhance background rejection capabilities.

\begin{figure}[h]
\begin{center}
\includegraphics[width=0.5\textwidth]{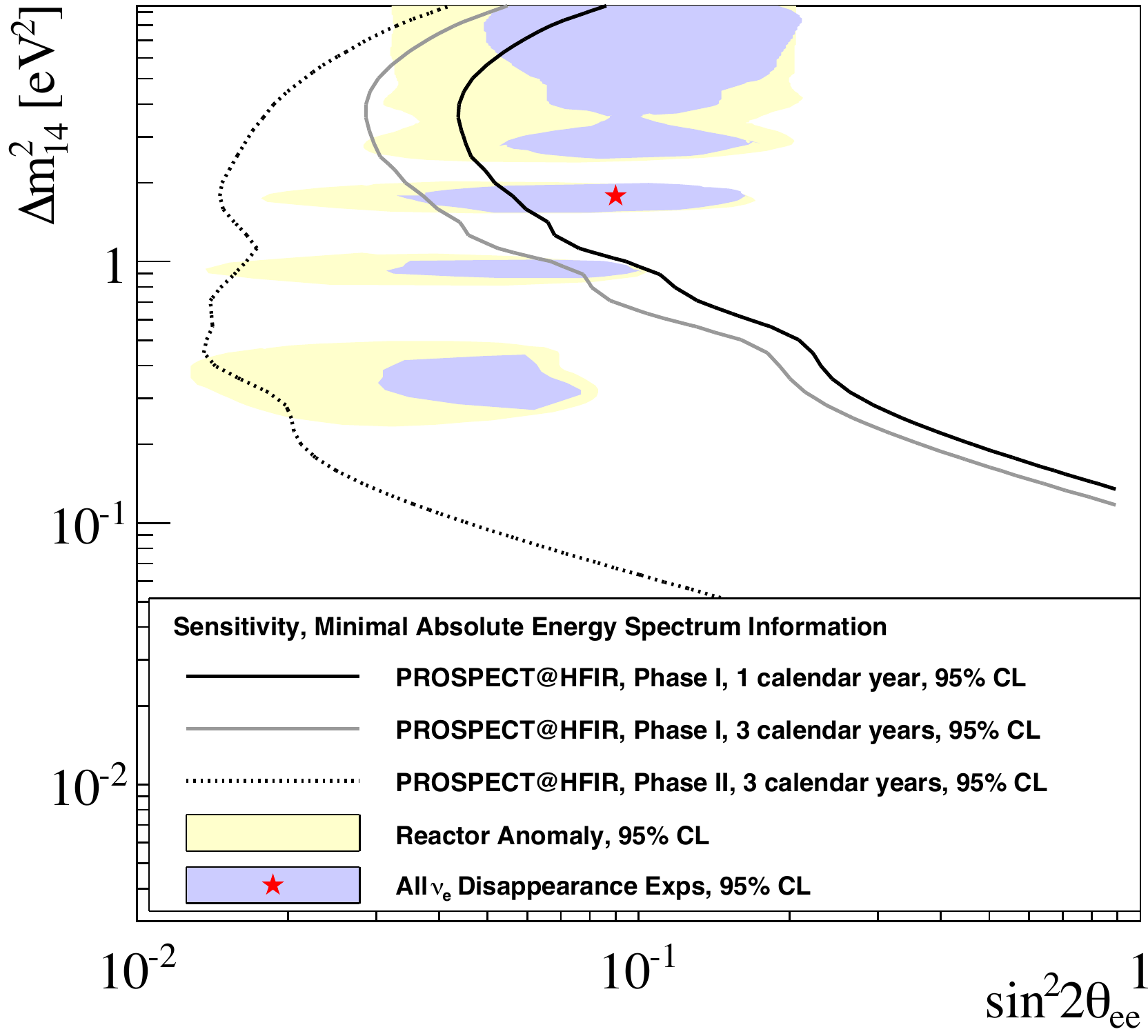}
\caption{The $3\sigma$ sensitivity curves for PROSPECT phase-1 and phase-2 operation. The red star indicates the sterile neutrino best-fit value from reactor and source experiments.}
\label{fig:PROSPECTsens}
\end{center}
\end{figure}

During operation at HFIR, PROSPECT will observe $\sim\!\!10^{3}$ IBD events per day. 
Fig~\ref{fig:PROSPECTsens} shows the sensitivity of both phase-1 and phase-2. 
Phase-1 is able to cover of the sterile neutrino preferred parameter space at $3\sigma$ within one calendar year and $5\sigma$ in three years.

\section{Background characterization}\label{sec-background}
For an experiment operating directly next to a reactor, background suppression is of vital importance.
This can be achieved through passive and active means via shielding and enhanced detector technology. 
To understand the background environment at HFIR, multiple detectors have been operated measuring gamma, muon, and fast and thermal neutron fields. 
Backgrounds can be split into those which are reactor-related (high energy gamma rays and thermal neutrons) and those that are inherent to the environment (cosmogenic showers and natural radioactivity). 

\subsection{Reactor-related backgrounds}
\begin{figure}[h]
\begin{center}
\includegraphics[width=.7\textwidth]{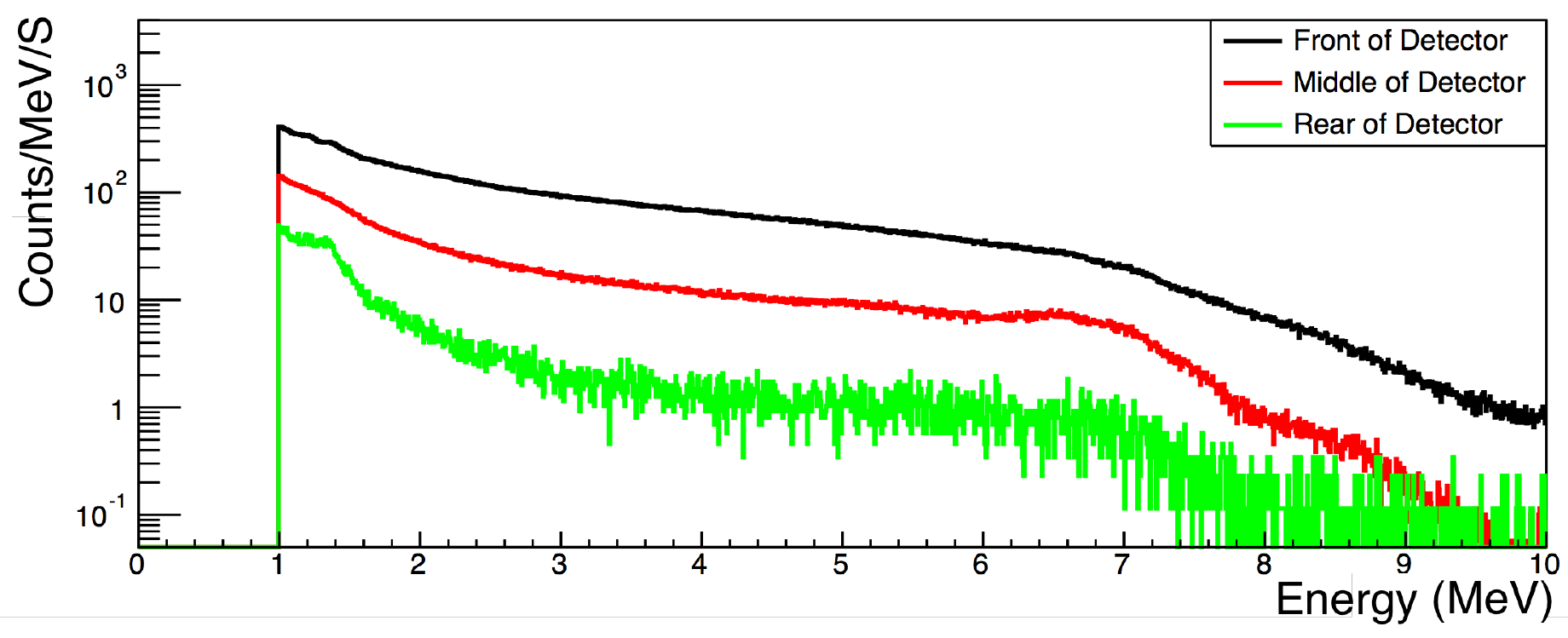}
\caption{The gamma response of an NaI detector placed at the front, middle, and rear of the phase-1 detector location at HFIR. There is a strong radial dependence of the gamma flux, which can be addressed with local shielding.}
\label{fig:HFIRGamma}
\end{center}
\end{figure}

To determine the gamma backgrounds at the near detector site, NaI and HPGE detectors were deployed at multiple locations, mapping out the gamma field as a function of distance from the reactor. 
Fig~\ref{fig:HFIRGamma} shows the gamma response of the NaI detector at the front, middle, and rear of the phase-1 detector site, approximately 7~m, 9~m, and 11~m from the rector core respectively. 

There is a strong reduction of the gamma flux at larger distances from the reactor. 
This allows for the design and deployment of targeted shielding to maximize the total active volume of the detector package.

The thermal neutron field was also measured and sources of thermal neutrons have been identified. 
These can be shielded with boron-loaded silicone rubber to minimize the production of high energy gammas from neutron capture on high-Z material.   

\subsection{Non-reactor related backgrounds}

The other main source of backgrounds for PROSPECT are cosmogenic muons, neutrons, and protons which can punch through passive shielding and mimic inverse-beta decay events.
PROSPECT will operate at only a few meter-water-equivalent of overburden, and will therefore have minimal shielding from cosmogenic activity.
To determine these backgrounds, both muon and fast neutron detectors were operated at multiple positions near the reactor. 

A noticeable reduction of the muon flux was observed from the reactor water pool, which shadows the detector site. 
The FaNS-1 fast neutron spectrometer~\cite{langford2014fast} was deployed at four sites near the detector location. 
The measured spectral shape from 1-200~MeV is consistent with that predicted by Ref~\cite{Gordon2004}. 

These measurements are part of a broad effort to fully characterize the environmental backgrounds in which PROSPECT will operate. 
The results can then be used to optimize shielding package design.  

\section{Lithium-loaded liquid scintillator}\label{sec-lils}
Apart from passive shielding, new detector technology is able to greatly reduce the effects of background radiation. 
Reactor antineutrino experiments typically use liquid organic scintillator as the detection medium. 
Antineutrinos interact with protons in the scintillator via inverse-beta decay (IBD), producing a positron and an epithermal neutron ($\left<E_n\right>=10$~keV).
By capturing the resulting neutron, a coincidence can be made to reduce backgrounds. 

\subsection{Neutron tagging}
Common neutron capture agents gadolinium and boron produce gamma rays in the final state, which can escape from smaller detectors.
Therefore, much work has been done to develop lithium-loaded liquid scintillator (LiLS) for capture-gated neutron spectroscopy~\cite{Fisher2011126, Bass:2012ur, Zaitseva2013747} and antineutrino detection~\cite{bugey:1996}.
Because neutron capture on lithium produces only charged particles in the final state ($n + Li \rightarrow \alpha + t + 0.5$~MeV$_{visible}$), the capture produces a consistent signal and does not suffer geometric effects. 
For application to neutrino detection, this provides a highly localized event, which can be used to precisely determine the location of an IBD event and also reject accidental coincidences. 

\begin{figure}[h]
\begin{center}
\includegraphics[width=.5\textwidth]{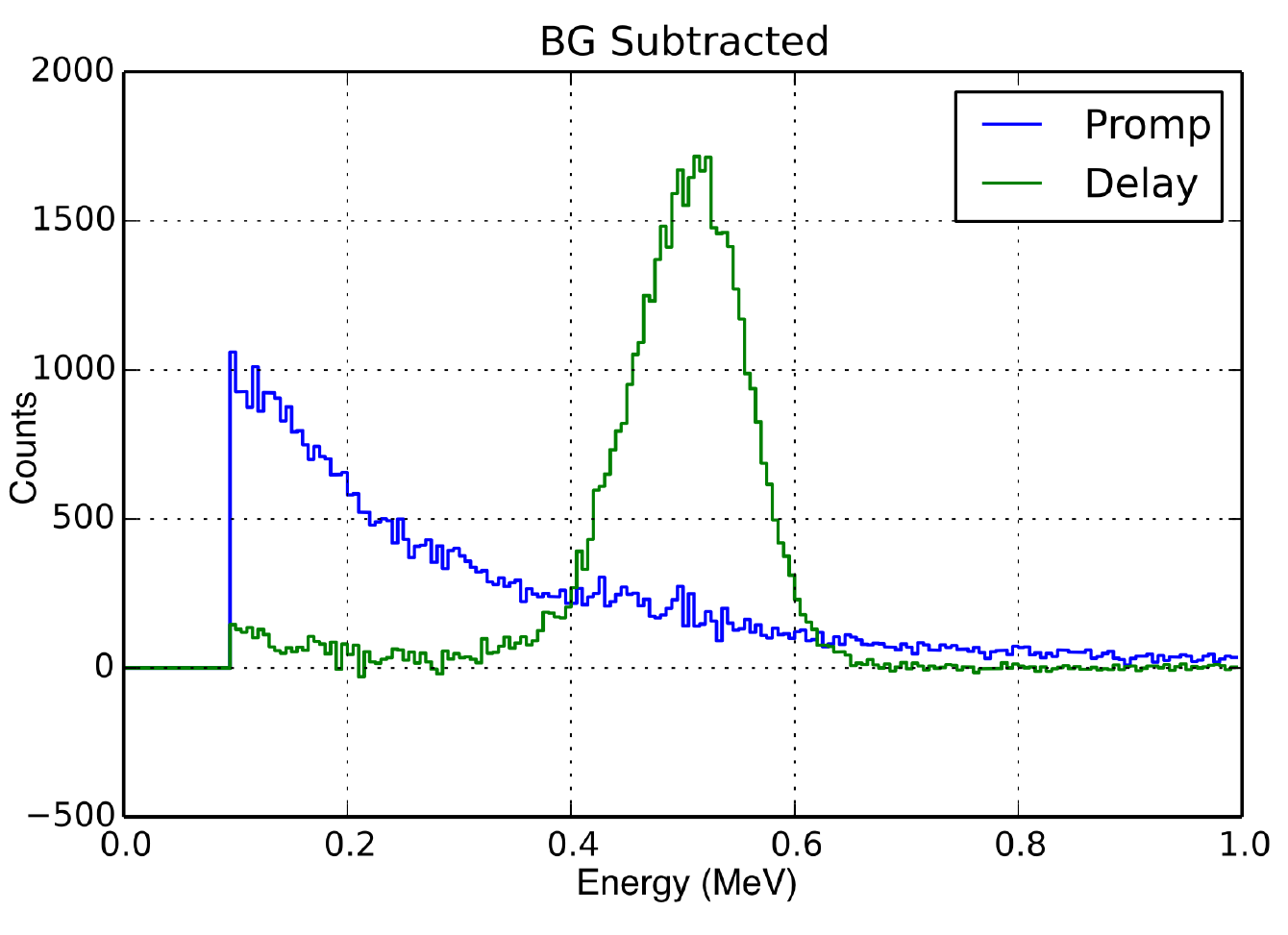}
\caption{The energy spectra from prompt and delay signals in a LiLS developed for PROSPECT. The prompt signals are fast neutron interactions and the delay signals are from neutron captures on lithium.}
\label{fig:LiLSnCapEn}
\end{center}
\end{figure}

For PROSPECT two new lithium-loaded scintillators have been developed based on two common scintillating media, linear alkyl benzene (LAB) and di-isopropyl naphthalene (DIN).
Both formulae have been characterized in 13~cm right cylinders readout by two ET9823\footnote{Certain trade names and company products are mentioned in the text or identified in illustrations in order to adequately specify the experimental procedure and equipment used. In no case does such identification imply recommendation or endorsement by the National Institute of Standards and Technology, nor does it imply that the products are necessarily the best available for the purpose.} photomultiplier tubes using a low-activity $^{252}$Cf source at Yale.
The PMTs are recorded using a CAEN V1720 digitizer with 250~MS/s sample rate and 12bit resolution. 
Fig~\ref{fig:LiLSnCapEn} shows the prompt (fast neutron interactions) and delay (neutron capture on lithium) energy spectra detected with the LAB-based sample. 

The observed neutron capture peak is well separated from noise and has a full-width half-max of approximately 15\%.  

\subsection{2D Pulse shape discrimination}
A second feature of liquid scintillator is the ability to perform pulse shape discrimination (PSD) between neutron and gamma interactions. 
This can be used on both the prompt and delay signals to identify IBD events, which have a PSD signature distinct from the two most common backgrounds (accidental coincidences between two gamma rays, and correlated fast neutron interactions).
An IBD event has a prompt signal that is "gamma-like" and a delay signal that is "neutron-like", while fast neutron interactions have "neutron-like" prompt and delay and accidental coincidences have "gamma-like" prompt and delay.
The power of this discrimination can be demonstrated by placing successive cuts on data. 
With preliminary cut values, a factor of 500 reduction is observed between the no-cut data and post IBD-like cuts. 


\section{Conclusions}
PROSPECT is a new experimental effort to search for short-baseline antineutrino phenomena, including sterile neutrinos and deviations from the predicted energy spectral shape.  
Work has been done to fully characterize the backgrounds at HFIR to further the optimization of shielding design and simulation.
Novel lithium-loaded liquid scintillator has been developed to enhance neutron tagging and allow for PSD to separate IBD events from fast neutron or accidental coincidences.
PROSPECT expects to observe  $\sim\!\!\!10^{3}$ IBD antineutrino events per day during reactor operation. 
As shown in Fig~\ref{fig:PROSPECTsens}, this will provide a ``prediction-free'' coverage to the sterile neutrino preferred parameter space in one year at $3\sigma$. 

\section{Acknowledgments}
This material is based upon work supported by the U.S. Department of Energy Office of Science. 
Additional support for this work is provided by Yale University. 
We gratefully acknowledge the support and hospitality of the High Flux Isotope Reactor and the Physics Division at Oak Ridge National Laboratory, managed by UT Battelle for the U.S. Department of Energy. 




\nocite{*}
\bibliographystyle{abbrv}
\bibliography{NOW2014}







\end{document}